\documentclass[twocolumn,twoside,slac]{revtex4}
\usepackage{graphicx}
\usepackage{fancyhdr}
\usepackage{here}
\begin{document}

\title{Development of a PCI Based Data Acquisition Platform for High Intensity
Accelerator Experiments}

%

\author{T.~Higuchi, H.~Fujii, M.~Ikeno, Y.~Igarashi, E.~Inoue,
R.~Itoh, H.~Kodama, T.~Murakami, M.~Nakao, K.~Nakayoshi, M.~Saitoh, S.~Shimazaki,
S.~Y.~Suzuki, M.~Tanaka, K.~Tauchi, M.~Yamauchi, Y.~Yasu}
\affiliation{High Energy Accelerator Research Organization (KEK), Tsukuba}

\author{G.~Varner}
\affiliation{University of Hawaii, Honolulu, Hawaii}

\author{Y.~Nagasaka}
\affiliation{Hiroshima Institute of Technology, Hiroshima}

\author{T.~Katayama, K.~Watanabe}
\affiliation{Densan Co. Ltd., Tokyo}

\author{M.~Ishizuka, S.~Onozawa, C.~J.~Li}
\affiliation{Designtech Co. Ltd., Tokyo}

\begin{abstract}
Data logging at an upgraded KEKB accelerator or the J-PARC facility, currently under commission,
requires a high density data acquisition platform with integrated data reduction CPUs.
To follow market trends, we have developed a DAQ platform based on the PCI bus, a choice
which permits a fast DAQ and a long expected lifetime of the system.
The platform is a 9U-VME motherboard consisting of four slots for signal
digitization modules, readout FIFOs for data buffering, and three PMC slots, on one of
which resides a data reduction CPU.
We have performed long term and thermal stability tests.
The readout speed on the platform has been measured up to 125 MB/s in DMA mode.

\end{abstract}

\maketitle

\thispagestyle{fancy}


\section{INTRODUCTION}
The Belle collaboration~\cite{bib:Belle} has established
the validity of the Kobayashi-Maskawa scheme of $CP$ violation
and has measured various quantities in $B$ meson decays~\cite{bib:CPV}.
These measurements have been facilitated by the KEKB accelerator~\cite{bib:KEKB},
which is an $e^+e^-$ collider with a remarkable
luminosity greater than ${\cal L} > 10^{34}$~cm$^{-2}$s$^{-1}$.
To extend the physics coverage to precise determination
of the CKM matrix elements, as well as a search for new physics,
we are planning a significant upgrade of the experiment for 
operating at a luminosity of ${\cal L} > 10^{35}$~cm$^{-2}$s$^{-1}$.

Concurrently, planning for experiments at a high intensity proton beam facility, 
J-PARC, which utilizes fixed target~\cite{bib:J-PARC} collisions, is maturing.
This experimental complex enables us to explore physics topics such
as neutrino mass
difference or $CP$ violation,
as well as a precise determination of $\epsilon'/\epsilon$
in $K$ meson decays.

In the high-luminosity experiments
the level-1 trigger rate and the data size per event
will increase to $\sim10$~kHz and 50-100~kB, respectively, and an overall requirement
for the data acquisition system (DAQ) is typically estimated as 500-1000~MB/s.
To match these boundary conditions, several DAQ components need to be developed:
triggers, timing distributors, DAQ platforms, computer farms, etc.

\subsection{Overview of Current Belle DAQ}
\begin{figure*}[t]
\centering
\resizebox{0.70\textwidth}{!}{\includegraphics{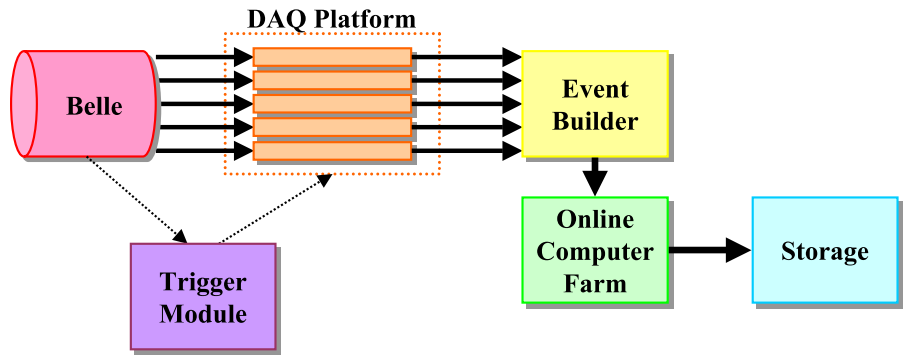}}
\caption{An illustration of the global data flow of the current Belle DAQ system.}\label{fig:global-dataflow}
\end{figure*}

Figure\ \ref{fig:global-dataflow} illustrates global data flow through the current Belle 
DAQ system~\cite{bib:DAQ}.
Hit signals from the Belle detector  
are transferred to electronics hut
via differential or other cables.
In the electronics hut,
hit signals are converted from analog to digital at the DAQ platform
and are stored in an event buffer to minimize dead-time.
These data are then fetched by CPUs
and experience pedestal subtraction, data suppression, and
data formatting.  The formatted data are sent to
an event builder to be combined into a single event record.
Output from the
event builder is then transferred to the online computer farm.  At the online
computer farm, data formatting, calibration, event reconstruction,
and event filtering, known collectively as level-3 trigger processing, is performed.
As the final step, the online computer farm transfers the data
to a mass storage system.

A level-1 hardware trigger module is drawn in Figure.\ \ref{fig:global-dataflow}.
Each sub-detector generates trigger signals (known as level-0 triggers), which
are fed to global decision logic, which determines whether the event should
be acted upon or discarded.  This trigger decision is first fed to a sequence
controller that ensures a correct timing handshaking by detecting
``busy'' signals from each DAQ platform.  The sequence controller
routes the trigger signal to the trigger timing distributor,
which then fans-out trigger signals to the DAQ platforms to initiate
the data fetch by the CPUs.

\subsection{Design of a New DAQ Platform}
To minimize bandwidth from the detector, data size should be
reduced as much as possible.  There are two major trends of data reduction:
one is complex triggering at lower levels, while the other
is a data reduction by online processors.
To save limited human resources, we are encouraged to use commercial products.
From the view point of the market trends, ASICs (Application-Specific Integrated Circuits)
are becoming very popular;
however, an evolutionary storm in the computer market is also brewing.
Considering these matters, we have decided to develop a new DAQ
platform equipped with CPUs for data compression.

The internal bus of the platform should be fast.  The PCI bus
can perform up to 133~MB/s, which is faster than the VME bus, upon which
the current Belle DAQ system is based.
The PCI bus also has an advantage in compactness and modularity
by providing a mezzanine card standard, known as PMC -- PCI mezzanine card.
Since the data size is expected to triple, a smaller unit platform is 
preferred so as to keep the space requirements within the electronics hut manageable.
The modularity of the PMC is especially beneficial when considering upgrade of the processor PMC module.
Moreover, because the PCI bus has spread so widely throughout the world,
it is unlikely to die quickly and current new technologies show little promise to improve on it.
Thus we can aim for a long lifetime of the PCI-based DAQ platform,
while using newest products.

Higher luminosity consequently requires a pipelined DAQ system.
It is necessary for the platform to be equipped with readout FIFOs to buffer events.

In summary, we have chosen an on-board DAQ platform based on the PCI bus~\cite{bib:PCI},
which contains a CPU for data compression and readout FIFOs for event buffering.

\section{COMMON DAQ PLATFORM}
The new DAQ platform consists of two components, the COPPER (COmmon Pipelined Platform for
Electronics Readout) and the SPIGOT (Sequencing Pipeline Interface with Global Oversight and Timing).
The COPPER is a 9U-VME sized card which serves as the platform for interfacing the front-end specific 
electronics, via a data compression CPU, to the event builder.  Detector input connections are made through
FINESSE (Front-end INstrumentation
Entity for Sub-detector Specific Electronics) A/D conversion modules, which transfer their data to readout 
FIFOs instrumented on the COPPER board.  
The SPIGOT, which interfaces with the COPPER at its backplane (rear) side,
is a 6U-VME sized board which includes a trigger module that issues trigger decisions to the FINESSE daughter-cards
on the COPPER.  In the following subsections, we give description of the COPPER and the SPIGOT.

\subsection{COPPER}
\begin{figure*}[t]
\begin{center}
\resizebox{0.70\textwidth}{!}{\includegraphics{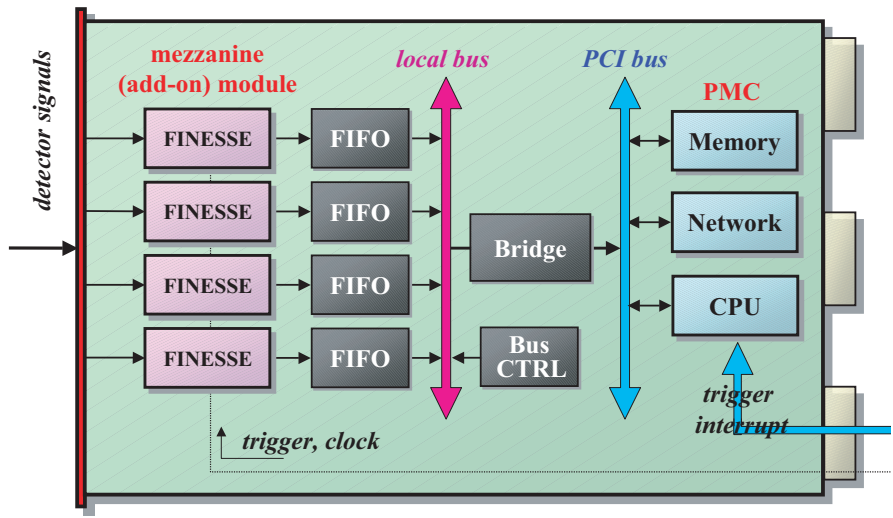}}
\end{center}
\caption{A block diagram of the COPPER board.}\label{fig:copper-block}
\end{figure*}
\begin{figure*}[t]
\begin{center}
\resizebox{0.70\textwidth}{!}{\includegraphics{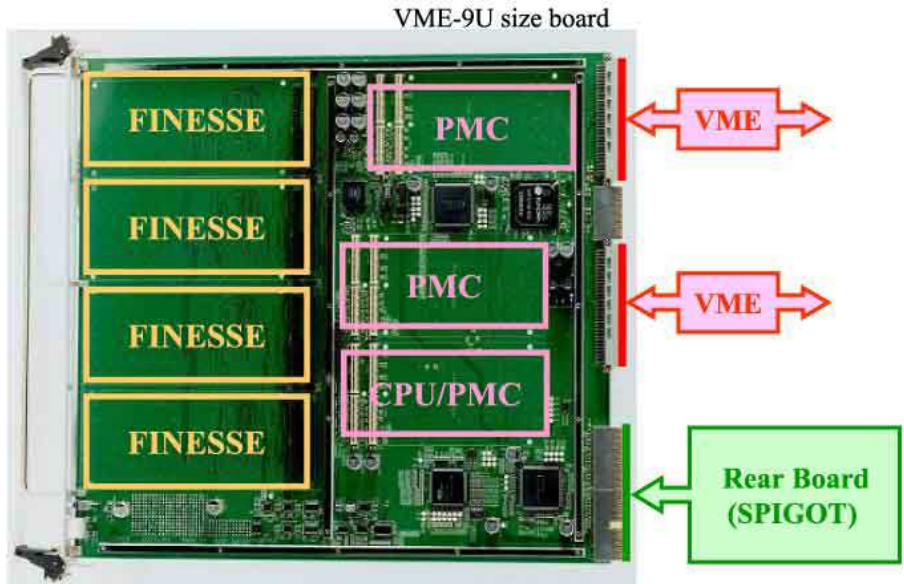}}
\end{center}
\caption{A photograph of the COPPER board, with various functional components and 
interfaces indicated.}\label{fig:copper-photo}
\end{figure*}
Figure\ \ref{fig:copper-block} shows a block diagram of a single COPPER board.
The COPPER is a 9U-VME sized board.
It is equipped with four slots for FINESSE modules,
readout FIFOs, and three PMC slots connected to the PCI bus.
A processor module for the data compression resides at one of three PMC slots.

Detector hit signals are first put into one of (up to) four FINESSE cards,
at which the signals are digitized.
Upon receipt of a level-1 trigger by the FINESSE from a trigger module on the SPIGOT,
the FINESSE pushes data into the readout FIFOs.
Each readout FIFO is connected to the local bus on the COPPER.  A local-PCI bus bridge, PCI9054~\cite{bib:PLX9054}, collects
data from the FIFOs and transfers them to the main memory in the processor module on the PCI bus using DMA.

A local bus sequencer on the local bus counts the number of words in each readout FIFO
associated to every trigger
to provide the event size to the processor module and to throttle the level-1 trigger when the FIFOs get full.

In the processor module, complicated data compression is performed:
pedestal subtraction, zero suppression, and data formatting.
Formatted data is subsequently transferred from the processor
via an external link device to the event builder.

The COPPER has RJ45 port so that the processor module
on the COPPER can be operated through the network.
The COPPER also has a VME-PCI dual port memory to be redundantly operated from the VME master module
and to export the event data at an very early stage of the system development.

The photograph of the COPPER is shown in Figure\ \ref{fig:copper-photo}.

\subsection{SPIGOT}
The SPIGOT is a 6U-VME sized board upon which is mounted a trigger module and up to two PMC modules,
where one of two is intended to be used for PMC link device, typically an Ethernet card,
connected to the event builder.
The SPIGOT resides behind the COPPER, mounted in a transition module location, offset to align at
the bottom of the P3 plane location.  
The PCI bus on the SPIGOT is electrically connected to the one on the COPPER via PCI-PCI bridge on the COPPER.

The trigger module distributes a trigger timing and sampling clock to each FINESSE module,
the signals of which come from an upstream transmitter module.
The trigger module includes a trigger FIFO, which is accessible to the processor via the PCI bus,
to hold an event counter and a trigger type.  This trigger information
is included in a header of each event by the processor.

\section{FINESSE}
\begin{figure*}[t]
\begin{center}
\resizebox{0.70\textwidth}{!}{\includegraphics{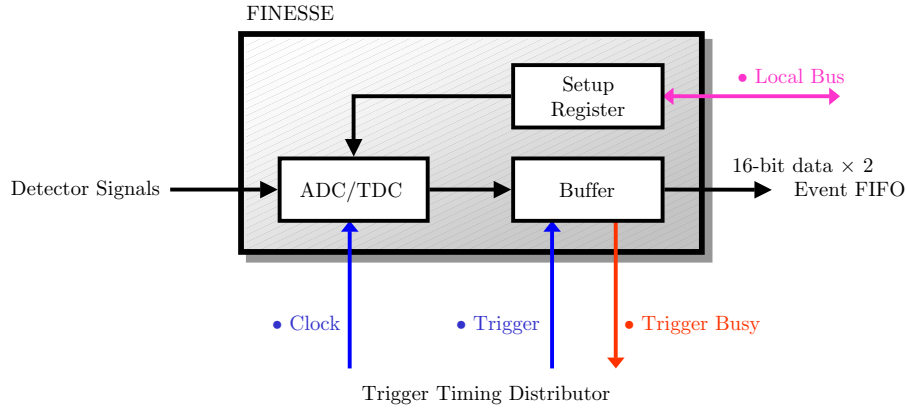}}
\end{center}
\caption{A block diagram of the FINESSE.  Digitized signals from the ADCs or TDCs
are stored in a synchronous buffer to minimize dead-time due to trigger latency.}\label{fig:finesse-block}
\end{figure*}
\begin{figure}[t]
\begin{center}
\resizebox{0.40\textwidth}{!}{\includegraphics{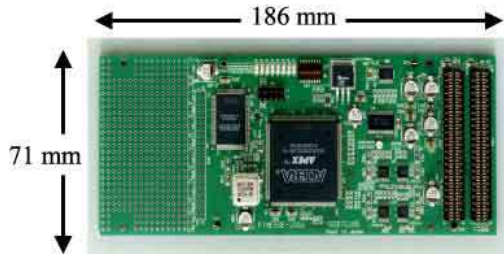}}
\end{center}
\caption{A photograph of a prototype FINESSE.}\label{fig:finesse-photo}
\end{figure}

The FINESSE is a mezzanine card with dimensions of $71.0\times186.0~{\rm mm}^2$ and is 
mounted on the COPPER.  Each FINESSE samples and digitizes detector hit signals according 
to either an external or internal clock,
holds digitized signals in its synchronous buffer,
and transfers the digitized data to readout FIFOs on the COPPER upon receipt of a level-1 trigger signal.
A block diagram of the FINESSE is shown in Figure\ \ref{fig:finesse-block}.  A photograph of a prototype 
FINESSE is shown in Figure\ \ref{fig:finesse-photo}.

The FINESSE receives an external sampling clock for A/D conversion continuously from the trigger module in 
LVDS format, and stores the digitized signals to the synchronous buffer according to this clock.
Once the FINESSE receives a trigger, it freezes the relevant contents of the synchronous buffer
moves these contents to a readout FIFO on the COPPER.

The FINESSE shall ensure a {\it busy} handshaking with the trigger module.
Upon receipt of a level-1 trigger, each FINESSE shall assert trigger busy until all data associated
with the trigger are flushed to the readout FIFO.
Delivery of the next trigger by the trigger module is suspended until
all {\it busy} signals asserted by each of four FINESSEs are cleared.

A/D conversion can be as accurate as 16 bits.
To utilize full PCI bandwidth, two data words are fetched by the processor in parallel simultaneously.
The first 16-bit data word appears in the lower 16-bits of the local data bus
and the trailing 16-bit word does in the higher.

Write clocks to the readout FIFOs during {\it busy} assertion are spied upon by a local bus sequencer
to count the number of stored words in the readout FIFO.

\section{PMC PROCESSOR MODULE}
\begin{figure}
\begin{center}
\resizebox{0.40\textwidth}{!}{\includegraphics{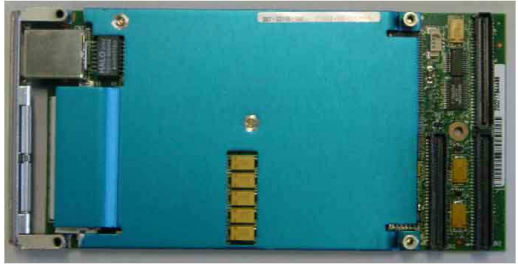}}
\end{center}
\caption{A photograph of the PMC processor board, a RadiSys EPC-6315.}\label{fig:EPC-6315}
\end{figure}
\begin{figure*}
\begin{center}
\resizebox{0.80\textwidth}{!}{\includegraphics{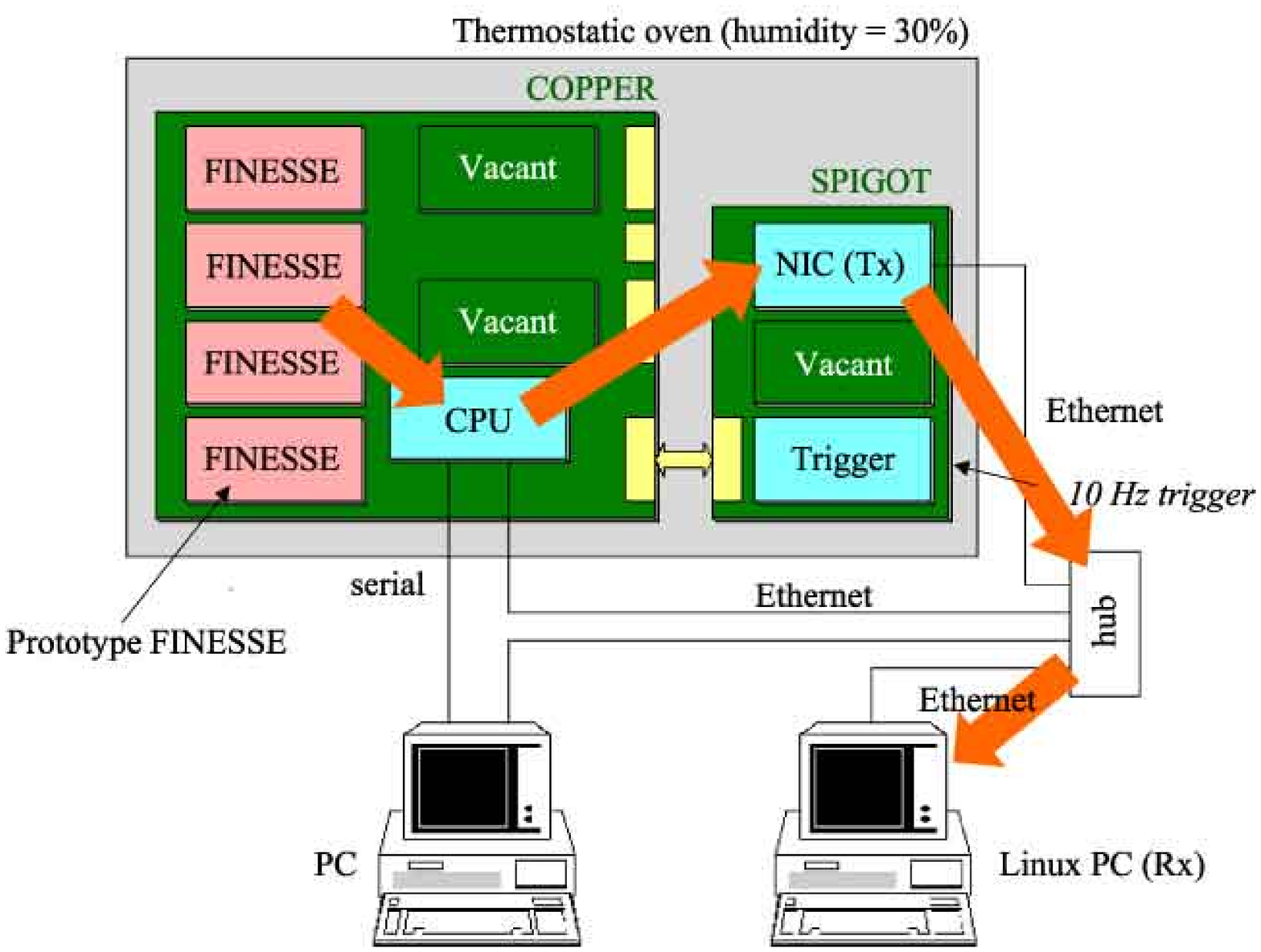}}
\end{center}
\caption{A schematic view of the test setup for the new DAQ platform.
Orange arrows show the data flow.
Data generated on the prototype FINESSEs are read by the processor module
(a RadiSys EPC-6315) on the COPPER.
After a data integrity check by the processor, the data are transferred
to the data receiver process running on a Linux PC through a network
interface module on the SPIGOT.
A 10~Hz random trigger is generated by the trigger module
on the SPIGOT and fed to the prototype FINESSEs.
Messages from the processor module are monitored on a serial terminal.
}\label{fig:test-setup}
\end{figure*}

The PMC processor module on the COPPER receives data from the readout FIFOs,
reduces the data size, and sends this formatted data to the event builder.

To develop the data reduction software easily, the processor should be capable
of handling a familiar operating system such as Linux.
It is also important for us to choose a CPU that is on a concrete upgrade path.
Our answer for the processor module is to use Intel processor based on a PC architecture.

Another constraint comes from the power consumption.
Because of the limited power supply capacity in a standard VME crate and because of
the limits imposed by air cooling, the power consumption by the processor module
should be less than 20~W.

We purchased EPC-6315 CPUs provided by RadiSys Corp.~\cite{bib:EPC6315}, 
equipped with Intel Pentium III processors operating with an 800~MHz clock.
In this configuration the processor consumes up to 17~W of electrical power.
Figure\ \ref{fig:EPC-6315} is a photograph of the EPC-6315.

We evaluated the performance of the EPC-6315 by means of a pseudo-data-compression software.
We assume that each FINESSE converts 16-channels of analog signal into 8-bit digital signals.
If we set the sampling rate and sampling window as 100~MHz and 500~ns, respectively,
the required data processing rate by the processor should be more than 32~MB/s at a 10~kHz trigger rate.
The test software works with Monte-Carlo-generated hit signals.
It copies data from a ``virtual readout FIFO (memory)'' to a work area,
subtracts pedestals, suppresses channels without hits, and compresses data
into hit timing and induced charge.
At 20\% occupancy, the processing speed is estimated 94~MB/s, which is marginal for our application.

We are developing a network based boot-up mechanism to handle the large number of processors.

\section{PERFORMANCE STUDY}
We have studied long-term stability, thermal stability,
and readout performance of the new DAQ platform.

Figure \ref{fig:test-setup} shows a block diagram of the test setup.
Data are generated on the prototype FINESSEs
according to a pre-defined table
and are read by the processor module
(EPC-6315) on the COPPER.
After a data integrity test on the processor, the data are transferred
to the data receiver process running on the Linux PC through a network
interface module on the SPIGOT.
In the receiver process, event header/footer, event counter,
and the contents are checked.  The event counter 
should be same for the four FINESSE modules and increase by 1 for each trigger.
The data contents should be consistent with the pre-defined table.
A 10~Hz random trigger is generated by the trigger module
on the SPIGOT and fed to the prototype FINESSEs.
Messages from the processor module are monitored via a serial terminal.

We operated the DAQ platform in the room temperature for 48~hours and
in a thermostatic oven for 77~hours,
varying the temperature from 10$^\circ$C to 50$^\circ$C.
We detected no errors in the receiver process.
We conclude the platform works stably for extended periods
and is tolerant of changes in the thermal environment.

We measured the data transfer speed on the COPPER
from the readout FIFOs to the CPU main memory using DMA.
We determined a transfer rate of 125 MB/s without DMA setup overhead,
which corresponds to 94\% of full PCI-32 perfomance.
We will continue to evaluate the system performance.

\section{SUMMARY}
A new DAQ platform is mandated for high-rate experiments to operate at an upgraded KEKB 
and a new J-PARC accelerator.
To realize a long lifetime of a DAQ platform while utilizing the newest products,
we have designed a platform motherboard based on the PCI bus and with an on-board data compression CPU,
equipped with a moderate computational capability.
Signal digitization modules, to be housed on the motherboard,
and a trigger distribution scheme are also designed.
The platform is measured to be stable and has achieved DMA data transfer
speeds of 125~MB/s.

\begin{acknowledgements}
This work is supported by Grant-in-Aid for Scientific Research (C) and
Grant-in-Aid for Scientific
Research on Priority Areas (B) from the Japan Society for the Promotion
of Science and
the Ministry of Education, Culture, Sports, Science and Technology.
\end{acknowledgements}

\end{document}